\documentclass[12pt]{article}
\usepackage{epsfig}
\usepackage{a4,isolatin1}
\usepackage{amsmath,amsfonts,latexsym, amssymb}
\usepackage{theorem}
\theorembodyfont{\upshape}
\newtheorem{satz}{Theorem}[section]
\newtheorem{defi}[satz]{Definition}

\newtheorem{bem}[satz]{Remark}
\newtheorem{lemma}[satz]{Lemma}
\newtheorem{koro}[satz]{Corollary}

\newtheorem{assumption}[satz]{Assumption}

\newtheorem{conclusion}[satz]{Conclusion}
\newtheorem{ob}[satz]{Observation}

\newcommand{\mcal}{\mathcal}
\newcommand{\mbf}{\mathbf}

\newcommand{\tit}{\textit}

\newcommand{\C}{\mathbb{C}}

\newcommand{\R}{\mathbb{R}}

\newcommand{\bewende}{ \hfill $\Box $}

\begin{document}
\thispagestyle{empty}
\begin{center}
\vspace*{1.0cm}

{\LARGE{\bf Scaling Analysis and Renormalisation Group for General\\
 (Quantum) Many Body Systems in the Critical Regime}} 

\vskip 1.5cm

{\large {\bf Manfred Requardt }} 

\vskip 0.5 cm 

Institut f\"ur Theoretische Physik \\ 
Universit\"at G\"ottingen \\ 
Bunsenstrasse 9 \\ 
37073 G\"ottingen \quad Germany\\
(E-mail: requardt@theorie.physik.uni-goettingen.de)

\end{center}

\vspace{1 cm}

\begin{abstract}
  With the help of a smooth scaling and coarse-graining approach of
  observables, developed recently by us in the context of so-called
  fluctuation operators (inspired by prior work of Verbeure et al) we
  perform a rigorous renormalisation group analysis of the critical
  regime. The approach is quite general, encompassing classical,
  quantum, discrete and continuous systems, the main thrust going to
  quantum many body systems. Our central topic is the analysis of the
  emergent properties of critical systems on the intermediate scales
  and in the scaling limit. To mention some particularly interesting
  points, we show that systems typically loose part of their quantum
  character in the scaling limit (vanishing of commutators) and we
  rigorously prove, with the help of the KMS-condition, the emergence
  of the phenomenon of critical slowing down together with the
  necessity of renormalising the time variable. These general features
  are then illustrated with the help of an instructive class of models
  and are related to the singular structure of quasi particle
  excitation modes for vanishing energy-momentum.
\end{abstract} \newpage

\section{Introduction}One of the central ideas of the
renormalization group analysis of, for example, the critical regime, is
\tit{scale invariance} of the system in the \tit{scaling limit}. This
is the famous \tit{scaling hypothesis} (as to the underlying working
philosophy compare any good text book of the subject matter like e.g.
\cite{Ma} and references therein). Central in this approach is the
so-called \tit{blockspin transformation}, \cite{Ka}. That is,
observables are averaged and appropriately renormalized over blocks of
increasing size. At each intermediate scale a new \tit{effective
  theory} is constructed and the art consists of choosing (or rather:
calculating) the \tit{critical scaling exponents}, so that the
sequence of effective theories converge to a (scale invariant) limit
theory, provided that the start theory lay on the \tit{critical
  submanifold} in the (in general infinite dimensional) parameter
space of theories or Hamiltonians.

We want to mention a slightly different approach to renormalisation
(see for example \cite{Amit}, \cite{Be} or \cite{Zinn}), which is more
in the spirit of renormalisation in quantum field theory. There exist
a lot of cross relations but in the following we do not discuss these
more technical aspects.  

Usually the calculations can only be performed in an approximative
way, the main tools being of a perturbative character and being
typically model dependent. Frequently, the discussionrelies on spin
systems to motivate and explain the calculational steps. While the
general working philosophy, based on the concepts of \tit{asymptotic
  scale invariance}, \tit{correlation length} and the like, is the
result of a deep physical analysis of the phenomena, there is, on the
other hand, no abundance of both rigorous \tit{and} model independent
results.

This applies in particular to the control of the convergence of the
scaled $l$-point correlation functions to their respective limits if
we start from a microscopic theory, lying on the \tit{critical
  submanifold}. In this case, correlations are typically long-ranged
and the usual heuristic arguments concerning the manipulations of
expressions containing \tit{block variables} of increasing size in the
face of long range correlations among the blocks becomes rather
obscure as one is usually cavalier as to the interchange of various
limit procedures.  One knows from examples, that this may be dangerous
in such a context.

Furthermore, the clustering of the higher correlation functions in the
various channels of phase space may be quite complex and non-uniform.
A concise and selfcontained discussion of the more general aspects and
problems, lurking in the background together with a useful series of
notes and references, can be found in \cite{Pa}, section 7.

Usually, the crucial scaling relation (the \tit{scaling hypothesis})
\begin{equation}W_l^T(Lx_1,\ldots,Lx_l;\mu^*)=L^{-l\cdot n}\cdot
  L^{l\cdot\gamma}\cdot W_l^T(x_1,\ldots,x_l;\mu^*)\end{equation}
which is conjectured to hold at the fixed point (denoted by $\mu^*$ in
the parameter space), is the starting point (or physical input) of the
analysis. Here, $W_l^T$ denote the truncated $l$-point functions (see
below), $L$ is the diameter of the blocks, $Lx_i$ denote the
centers of the blocks, $n$ is the space dimension, $\gamma$ the
statistical renormalisation exponent. If it is different from $n/2$
or, rather, the expected naive value, we have an `\tit{anomalous}'
scale dimension (for convenience we have assumed all observables to scale
with  the same scale dimension). 

In the following analysis, one of our aims is a rigorous investigation
of such (and similar) scaling relations for the $l$-point functions,
starting from the underlying microscopic characteristics of the
theory. We will do this in a quite general manner, that is, the
underlying model theory can be \tit{classical} or \tit{quantum},
\tit{discrete} or \tit{continuous}. We try to make
only very few and transparent assumptions as it is our strategy, to
deal only with the really characteristic (almost model independent)
aspects of the subject matter. A central goal of our analysis is a
rigorous discussion of a number of characteristic properties of both the
\tit{intermediate} and \tit{limit states}, the observables and
dynamics occurring on these levels of renormalisation etc.,
with special emphasis on the quantum aspects.

In section 2 we develop the conceptual framework and a variety of
technical tools. As a technical side aspect we discuss the differences
between our \tit{smooth scaling} approach and the perhaps more common
averaging over sharp blocks. In section 3 we show that classical
continuum systems can be easily incorporated into our general scheme.
In section 4 we rigorously study a large class of models which can be
treated from a unified point of view. We exhibit the close connection
between the critical exponents and the spectral properties of the
correlation functions for vanishing energy-momentum. In section 5 we
analyse characteristic properties of the system on the intermediate
scales and in the scaling limit. Among other things we show that the
system may loose some of its quantum character in the scaling limit
(vanishing of commutators).  As a perhaps particularly interesting
result we provide a rigorous proof of the phenomenon called
\tit{critical slowing down} (based on the KMS-condition) together with
a renormalisation of the characteristic time scale of the dynamical
evolution (see section (\ref{Slowing})). We show, using the class of
models studied in section 4, that the dynamic scaling exponent,
occurring in the renormalisation of the time variable, is closely
related to the energy-dispersion law of some quasi-particle excitation
branch for vanishing momentum.

What regards the general working philosophy, one should perhaps
mention the framework, expounded in e.g. \cite{Bu} in the context of
the analysis of the \tit{ultraviolet behavior} in algebraic quantum
field theory, or, in the classical regime, the approach of e.g. Sinai
(\cite{Sinai}). While our framework also comprises the classical
regime (cf. the discussion concerning classical continuous systems in
section \ref{class}), it is mainly designed to deal with the more
complicated quantum case.  In so far, it is an extension of the
methods, developed by us in \cite{Re1}, which, on their side, were
inspired by prior work of Verbeure et al; see the corresponding
references in \cite{Re1}.  Recently we became aware of a nice
treatment of the block spin approach in the quantum regime in the bock
of Sewell (\cite{Sewell}), who employs methods which are different
from ours, but are complementing them (\tit{quantum (non-) central
  limits}).

As there exist presumably several thousand papersin this field, we
feel unable to relate our own approach to all the other approaches or
to make a detailed analysis of what is entirely new. Our main thrust
goes in the direction of a conceptual analysis and the developement of
a coherent and general point of view. In this respect we think, our
presentation is reasonably self-contained and contains a number of
original results. We briefly note that, in order to keep the paper
within reasonable length, we chose to perform most of the long and
intricate technical analysis of the scaling behavior of $l$-point
correlation functions on the critical submanifold elsewhere. A
preliminary treatment of this particular problem can be found in the
second part of \cite{0205}.

We end this introduction with mentioning a perhaps subtle point. In
the following we concentrate most of our analysis on the hierarchy of
correlation functions which can be used to define the theory. We
generate renormalized limit correlation functions from them which
happen to be scale invariant (in a sense clarified below), thus
defining a new limit theory via a \tit{reconstruction process}. On the
other hand, we do not openly discuss the flow of, say, the
renormalized Hamiltonians through parameter space as a sequence of
more and more \tit{coarse-grained} effective Hamiltonians. The
characteristics of these renormalised intermediate theories are
however implicitly given by their hierarchy of correlation functions
as was already explained in e.g. \cite{Re1} or \cite{Bu}. 

One should therefore emphasize, that this well-known integrating out
or decimation of degrees of freedom, which characterizes the ordinary
approaches is \tit{automatically} contained in our approach! The
effective time evolution is carried over from the microscopic theory
as described in \cite{Re1} or (in a slightly other context) in
\cite{Bu}, see also \cite{Verbeure} and is redefined on each
intermediate scale, thus implying automatically a rescaling or
renormalisation of both the time evolution and the corresponding
Hamiltonian; see section \ref{rigorous}. In case we work in an
scenario, defined by ordinary Gibbs states, our framework would
exactly yield these effective Hamiltonians. Nevertheless, it is an
interesting task, to apply our method directly to the microscopic
Hamiltonian.
\section{The Conceptual Framework}
\subsection{Concepts and Tools}
As to the general framework we refer the reader to \cite{Re1}. One of
our technical tools is a modified (smoothed) version of averaging
(modifications of the ordinary averaging procedure are also briefly
mentioned in the notes in \cite{Pa}). Instead of averaging over blocks
with a sharp cut off, we employ a smoothed averaging with smooth,
positive functions of the type
\begin{equation}f_R(x):= f(|x|/R)\quad\text{with}\quad f(s)=
\begin{cases}1 & \text{for $|x|\leq 1$} \\ 0 & \text{for $|x|\geq 2$}
\end{cases}
\end{equation}
Remark: We will see in the following, that the final result is more or
less independent of the particular class of averaging
functions!\\[0.3cm]
We note that this class of scaled functions has a much nicer behavior
under Fourier transformation, as, for example, functions with a sharp
cut off, the main reason being that the tails are now also scaled. We
have
\begin{equation}\hat{f}_R(k)=const\cdot R^n\cdot \hat{f}(R\cdot
  k)\end{equation}
One might perhaps think that this choice of averaging will
lead (as a consequence of the scaled tails) to a limit theory, being
different from one, constructed by employing a sharp cut-off. This is
however not the case. As the mathematical differences between the two
approaches, that is, using either sharp or smooth cut off
functions, are technically a little bit subtle and perhaps not so
apparent, we discuss some of the technical aspects below. 

We briefly describe the implications coming of
translation invariance. We have for the correlation functions
\begin{equation}W(x_1,\ldots,x_l)=W(x_1-x_2,\ldots,x_{l-1}-x_l)\end{equation}
The truncated correlation functions are defined inductively as follows
(see \cite{Re1})
\begin{equation}W(x_1,\ldots,x_l)=\sum_{part}\prod_{P_i}W^T(x_{i_1},\ldots,x_{i_k})\end{equation}
the sum running over all partitions of the set $\{x_1,\ldots,x_l\}$.
The (distributional) Fourier transform reads
\begin{equation}\tilde{W}^T(p_1,\ldots,p_l)=\hat{W}^T(p_1,p_1+p_2,\ldots,p_1+\cdots
p_{l-1})\cdot\delta(p_1+\cdots p_l)\end{equation}
The dual sets of variables are
\begin{equation}      y_i:= x_i-x_{i+1}\;,\;q_i=\sum_{j=1}^i p_j\quad i\leq
  (l-1)\end{equation}  

In contrast to the averaging procedure, introduced above, the usual
\tit{block-variable}-averaging is a \tit{sharp cutoff} averaging,
performed for example over balls, $B_R$, of radius $R$. That is,
observables are integrated over balls, $B_R$, with the help of the
incidence functions
\begin{equation}\chi_R(x):= \chi_1(|x|/R)\quad\text{with}\quad \chi_1(x)=
\begin{cases}1 & \text{for $|x|\leq 1$} \\ 0 & \text{for $|x|>1$}
\end{cases}
\end{equation}
The averaging over operators, leading to the so-called \tit{block} or
\tit{fluctuation operators}, is performed in the following way:
\begin{equation}A_R:=(const)\cdot R^{-\gamma}\cdot\int A(x)\cdot
  f_R(x)d^nx\end{equation}
 with the exponent $\gamma$ suitably chosen
and $const$ being a possible numerical and unimportant multiplicative
constant (related e.g. to the volume of the unit ball or something like
that). A corresponding expression holds for $\chi_R$ replacing $f_R$.
\\[0.3cm]
Remark: Here and in the following, the $A$'s are always normalized to
$\langle A\rangle=0$, in order to really get the pure fluctuation
effects.\vspace{0.3cm}

At this point we want to state a general principle which allows to
choose an appropriate scaling exponent, $\gamma$. As in most of the
discussions in the literature only a particular fixed field,
$\phi(x)$, or spin, $S(x_i)$, is employed, it is frequently not clear
that something has actually to be said in this context. This holds the
more so as quite a few different renormalisation schemes (or
philosophies) are used in practice, with the tacit understanding that
the \tit{critical exponents} are physically apriori given and
insensitive to the concrete decimation procedure. This problem
becomes, in our view, more virulent in the quantum regime with,
usually, a lot of different non-commuting observables.

We think that, if we adopt a more general viewpoint, the necessary
general principles become more apparent. This holds also for what we
call a possible \tit{problem of consistency}. This problem consists of
the following points.
\begin{ob}\hfill\\
\begin{enumerate}
\item It is reasonable to choose the scaling exponents so that certain
  two-point auto-correlation functions of block observables survive in
  the scaling limit. Note that an inappropriate choice drives the
  auto-correlation functions either to zero or infinity!
\item The Cauchy-Schwarz inequality then guarantees that at least the
  corresponding mixed two-point functions remain finite in the limit.
\item On the other hand, this shows that one may have a certain
  freedom in selecting the correlation functions and, by the same
  token, the observables one wants to survive in the limit. This will
  of course affect the structure of the possible limit theories.
\item If, on the other hand, we have a lot of different
  (non-commuting) quantum observables together with their composites,
  it is presumably not an easy task to make all these (possibly
  independent) choices in a consistent way so that a coherent Hilbert
  space structure results in the limit. This problem becomes virulent
  if we end up with a theory having non-vanishing higher truncated
  correlation functions. The reason is that possible obstructions may
  result from the decay behavior of higher n-point functions in the
  difference variables which has to be in complete balance with the
  chosen scaling exponents. These, on their side, are already fixed by
  the 2-point functions! We briefly discussed this issue in the last
  section of \cite{Re1} and we make a more detailed analysis in the
  second part of our paper \cite{0205}.
\end{enumerate}
\end{ob}
\begin{conclusion}We fix the
  renormalisation exponents, $\gamma_i$, of the respective observables
  via the non-vanishing and finiteness of (a class of) 2-point
  auto-correlation functions. This will yield constraints on the
  scaling behavior of higher correlation functions, the consistency of
  which we then can check.
\end{conclusion}

In the following we will mainly employ the smooth cut-off procedure
which leads to a more transparent behavior of various expressions in
Fourier-space. It is satisfying that in the cases, we can actually
control, it leads to results being identical to the version with sharp
volume cut-offs. In order to compare these two cut-off conventions we
study in a first step various peculiar properties of the averaging
functions, $\chi_R(x)$. The Fourier transform of the smooth functions,
$f_R(x)$, are again smooth, living in the Schwartz-space, $\mcal{S}$,
i.e., decrease fast together with all their derivatives. In
\cite{Re1} we crucially employed $L^1$ or $L^2$ properties of various
expressions. In contrast to $\hat{f}_R(k)$, the $\hat{\chi}_R(k)$'s are
no longer in $L^1$ as $\chi_R(x)$ has a jump discontinuity. On the
other hand, it is in $L^2$ as
\begin{equation}\infty>\int |\chi(x)_R|^2d^nx=\int
  |\hat{\chi}_R(k)|^2d^nk\end{equation}
We have the little lemma
\begin{lemma}The Fourier transform, $\hat{\chi}_R(k)$, is in
  $\mcal{C}^{\infty}\cap \mcal{C}_0$ but not in $L^1$. It is however
  in $L^2$. We have the same scaling behavior for $\hat{\chi}_R(k)$ as
  for $f_R(k)$, that is
\begin{equation}\hat{\chi}_R(k)=const\cdot R^n\cdot\hat{\chi}_1(R\cdot
  k)      \end{equation}
\end{lemma}
Proof: The first statement follows from the compact support of $\chi_R(x)$
and the Riemann-Lebesgue lemma. The second statement follows as in the
smooth case.\bewende

An explicit calculation for $n=3$ yields:
\begin{equation}\hat{\chi}_1(k)=const\cdot
  |k|^{-3}\cdot\int_0^{|k|}r\cdot\sin(r)dr \end{equation}
For $|k|\to 0$ the integral is proportional to $|k|^3$. Furthermore we
can show that the expression is in fact infinitely differentiable in
$|k|=0$. For $|k|\to\infty$ a partial integration yields an expression
proportional to $-|k|\cdot\cos(|k|)+\sin(|k|)$. That is, we have in
leading order for $|k|\to\infty$:
\begin{equation}\hat{\chi}_1(k)\sim |k|^{-2}\;\text{for}\;n=3  \end{equation}

We mention some further peculiar properties of the indicator function,
$\chi_B(x)$, not shown by other functions. From
$\chi_B(x)=\chi_B(x)\cdot\chi_B(x) $ we infer for the Fourier
transform
\begin{equation}\hat{\chi}_B(k)=\hat{\chi}_B\ast\hat{\chi}_B(k) \end{equation}
and correspondingly for higher powers.
\begin{koro}By Young`s inequality (see e.g. \cite{Reed2}) we know,
  that in general the convolution of $L^2$-functions is only in
  $L^{\infty}$. The preceding formula shows that the convolution of
  $\hat{\chi}_B(k)$ with itself is again in $L^2$.
\end{koro}
Note that such a result is not immediately evident from the concrete form
of the respective Fourier transforms. In the case $n=1$ say, the
Fourier transform is essentially of the form $\sin(k)/k$. The result
for the convolution comes about due to the peculiar oscillatory
character of the expression and would \tit{not} hold for
e.g. $|\hat{\chi}_1(k)|$. We will briefly analyse in the following
subsection to what extent the renormalisation process is influenced by
these slightly nasty features of sharp cut-off functions. 
\subsection{The case of Normal Fluctuations}
As in \cite{Re1}, we assume that away from the critical point the
truncated $l$-point functions are integrable, i.e. $\in
L^1(R^{n(l-1)})$, in the difference variables,\\ $y_i:=x_i-x_{i+1}$. As
observables we choose the translates
\begin{equation}A_R(a_1),\ldots,A_R(a_l)\;,\;A_R(a):=R^{-n/2}\cdot\int
A(x+a)f(x/R)d^nx\end{equation}
(where, for convenience, the labels $1\ldots l$ denote also possibly
different observables). We then get (for the calculational details see
\cite{Re1}, the hat denotes Fourier transform, translation invariance
is assumed throughout, the $const$ may change during the calculation
but contains only uninteresting numerical factors):
\begin{multline}\langle A_R(a_1)\cdots A_R(a_l)\rangle^T=const\cdot
  R^{ln/2}\cdot\\\int \hat{f}(Rp_1)\cdots \hat{f}(-R[p_1+\cdots
 + p_{l-1}])\cdot \hat{W}^T(p_1,\ldots,p_{l-1})\cdot
 e^{-i\sum_1^{l-1}p_ia_i}\cdot e^{ia_l\sum_1^{l-1}p_i}\prod dp_i\\
=const\cdot R^{ln/2}\cdot R^{-(l-1)n}\cdot\\\int \hat{f}(p_1')\cdots
\hat{f}(-[p_1'+\cdots+p_{l-1}'])\cdot
\hat{W}^T(p'_1/R,\ldots,p'_{l-1}/R)\cdot e^{-i\sum_1^{l-1}(p'_i/R)a_i}\cdot e^{ia_l\sum_1^{l-1}p'_i/R}\prod dp'_i\end{multline}
We now scale the $a_i$'s like
\begin{equation}a_i:=R\cdot X_i\;,\;X_i\;\text{fixed}\end{equation}
This yields
\begin{multline}\langle A_R(R\cdot X_1)\cdots A_R(R\cdot
  X_l)\rangle^T=\\const\cdot R^{(2-l)n/2}\cdot \int
  e^{-i\sum_1^{l-1}p'_iX_i}\cdot e^{iX_l\sum_1^{l-1}p'_i}\cdot\\
 \hat{f}(p_1')\cdots\hat{f}(-[p_1'+\cdots+p_{l-1}'])\cdot\hat{W}^T(p'_1/R,\ldots,p'_{l-1}/R)\prod dp'_i\end{multline}

As the $\hat{f}$ are of strong decrease and $\hat{W}^T$ continuous and
bounded by assumption ($W^T\in L^1(\R^{n(l-1)})$!), we can perform the
limit $R\to\infty$ under the integral (Lebesgues' theorem of dominated
convergence)
and get:\\[0.3cm]
Case 1 ($l\ge3$):
\begin{equation}\lim_{R\to\infty} \langle A_R(R\cdot X_1)\cdots A_R(R\cdot
  X_l)\rangle^T=0\end{equation}
Case 2 ($l=2$):
\begin{equation}\lim_{R\to\infty}\langle A_R(R\cdot X_1)A_R(R\cdot
  X_2)\rangle^T=const\cdot\int \hat{W}^T(0)\cdot
  e^{-ip'_1(X_1-X_2)}\cdot\hat{f}(p'_1)\cdot\hat{f}(-p'_1)dp'_1\end{equation}
We arrive at the conclusion
\begin{conclusion}Assuming $L^1$-clustering in the normal regime away
  from the critical point and employing a smooth cut-off, all the
  truncated correlation functions vanish in the limit $R\to\infty$
  apart from the $2$-point function.  We hence have a quasi free
  theory in the limit as described in \cite{Re1} or in the work of
  Verbeure et al (cf. the references in \cite{Verbeure})
\end{conclusion}

In the case of smooth averaging we employ the transparent behavior of
the Fourier transformed expressions. On the other hand, the Fourier
transform is inherently non-local, which sometimes makes the analysis
more complicated. When using instead the sharp cut-off convention, we
described above, the behavior of the respective Fourier transforms
becomes opaque in the general case. On the other hand, we can try to
stay in coordinate space and perform the analysis there. Proceeding as
in the smooth case but avoiding Fourier transformation we get after
some straightforward manipulations
\begin{multline}\langle
  A_R(RX_1)\cdots A_R(RX_l)\rangle^T=R^{-ln/2}\cdot R^n\cdot\int
  W_l^T(y_1',\ldots,y_{l-1}')\\\cdot\chi_1(\sum_{j=1}^{l-1}R^{-1}y'_j+x''_l-\sum_{j=1}^{l-1}Y_j-X_l)\cdots\chi_1(x_l''-X_l)dy'_1\ldots dx_l''\end{multline}
with $x_i-x_l=\sum_i^{l-1}y_j$ by the definition in the preceding
subsection and
\begin{equation}x'_i=x_i+RX_i\;,\;x'_i-x'_l=\sum_i^{l-1}y'_j\;,\;x_l'':=x'_l/R \end{equation}

Again, the limit can be performed under the integral and is zero for
$l>2$. For the two-point function we get
\begin{equation}\lim_{R\to\infty}\langle
  A_R(RX_1)\cdot A_R(RX_2)\rangle^T= \int W^T_2(y)dy\cdot\int
  \chi_1(x-Y)\chi_1(x)dx \end{equation}
with $Y:=X_1-X_2$. We hence conclude:
\begin{conclusion}In the normal situation of $L^1$-clustering of
  correlation functions and sharp cut-off functions,
  $\chi_R(x)=\chi_1(x/R)$, we get the same results as in the case of
  smooth cut-off functions. However, to prove this, we have to perform
  the analysis in real space and avoid Fourier transformation.
\end{conclusion} 
\begin{koro}It is obvious from the preceding discussion that the
  particular form of the averaging functions, $f_R(x)=f(x/R)$, need
  not even simulate a volume averaging. For the argument to hold, it
  is e.g. sufficient that $f(x)$ is bounded with $\hat{f}(0)\neq 0$ and has compact support.
  What only changes is an unimportant multiplicative factor, depending
  on the type of function, being chosen.
\end{koro}
\subsection{The Relation to the Heuristic Scaling\\ Hypothesis}
In the following sections we develop a rigorous approach to \tit{block
  -spin renormalisation} in the realm of quantum statistical
mechanics, which tries to implement the physically well-motivated but,
nevertheless, to some extent heuristic scaling hypothesis. The
analysis will be performed both in coordinate space and Fourier space.
In this subsection we restrict our discussion to the two-point
correlation function, for which the asymptotic behavior is simpler and
more transparent.\\[0.3cm]
Remark: In the rest of the paper we replace the exponent $n/2$ in the
definition of $A_R(a)$ by a scaling exponent $\gamma'$, which will usually be
fixed during or at the end of a calculation. It plays the role of a
\tit{critical scaling exponent} and is model dependent.\vspace{0.3cm}

Let us hence study the behavior of
\begin{multline}\label{two} \langle A_R(R\cdot X_1)A_R(R\cdot
  X_2)\rangle^T=R^{-2\gamma'}\cdot\int W^T((x_1-x_2)+R(X_1-X_2))\\\cdot
  f(x_1/R)f(x_2/R)dx_1dx_2\\
=R^{-2\gamma'+2n}\int W^T(R[(x_1-x_2)+(X_1-X_2)])\cdot
f(x_1)f(x_2)dx_1dx_2
\end{multline}
We make the physically well motivated assumption that, in the critical
regime, $W^T$ decays
asymptotically like some inverse power, i.e.
\begin{equation}W^T(x_1-x_2)\sim(const+ F(x_1-x_2))\cdot
  |x_1-x_2|^{-(n-\alpha)}\quad 0<\alpha<n\;,\; F(x)\in L^1\end{equation}
for $|x_1-x_2|\to\infty$, $F$ bounded and well-behaved.

From the last line of $(\ref{two})$ we see that, as $f$ has compact
support, we can replace $W^T$, for $(X_1-X_2)\neq 0$ and $R\to\infty$ 
by its asymptotic expression and get for $R$ large:
\begin{equation}\langle A_R(R\cdot X_1)A_R(R\cdot
  X_2)\rangle^T\approx const\cdot R^{-2\gamma'+2n}\cdot
  R^{-(n-\alpha)}\cdot\int |y+Y|^{-(n-\alpha)}\cdot f\ast f(y)dy\end{equation}
We choose now 
\begin{equation}\gamma'=(n+\alpha)/2\end{equation}
and get a limiting behavior (for $R\to\infty$) as
\begin{equation}const\cdot\int |y+Y|^{-(n-\alpha)}\cdot f\ast f(y)dy\end{equation}
with $y=x_1-x_2,Y=Y_1-Y_2$ and
\begin{equation}f\ast f(y):=\int f(y+x_2)\cdot f(x_2)dx_2\end{equation}

We see that in contrast to some of the general folklore, the limit
correlation functions are not automatically \tit{strictly} scale
invariant but still depend in the above integrated manner on the
chosen smearing functions, $f$. Full scale invariance is recovered in
the regime $Y\to\infty$. Central in the renormalisation group idea is
that systems on the \tit{critical surface} (i.e., critical systems)
are driven towards a \tit{fixed point}, representing a scale invariant
theory.  This idea is usually formulated in an abstract parameter
space of, for instance, Hamiltonians. In our correlation function approach
scale invariance at the presumed fixed point would prove its existence
via the scaling properties of the correlation functions, that is
\begin{equation}W^T_2(L\cdot(X-Y);\mu^*)=L^{-2(n-\gamma')}W_2^T(X-Y;\mu^*)\end{equation}
with $\mu^*$ describing the fixed point in the (usually) infinite
dimensional parameter space. We see from the above that this picture
is \tit{asymptotically} implemented by our above limiting correlation
functions, as we have (with the choice $\gamma=(n+\alpha)/2$):
\begin{equation}W_2^T(X-Y;\mu^*)\sim|X-Y|^{-(n-\alpha)}\end{equation}
in the asymptotic regime. That is, the above scaling limit leads to a
limit (i.e. fixed point) theory, reproducing the asymptotic
behavior of the original (microscopic) theory.

One should however note that in the more general situation of
$l$-point correlation functions we have to expect a more complex decay
behavior and the existence of various channels as varying clusters of
observables move to infinity.
\subsection{Strategies for a Renormalisation Analysis on the Critical
  Surface}
Typically, the numerical scaling analysis is developed for the system
being away from the critical surface. The reason is that away from
criticality, under the heuristic assumption of e.g. exponential
clustering, the analysis is not beset with technical difficulties as,
for example, the interchange of limits and dealing correctly with long
range tales in correlation functions. It is then frequently argued
that, in case the system is sufficiently near to the critical surface,
the orbits of renormalized model systems nevertheless will approach
the vicinity of the fixed point, so that one can make a \tit{linear}
stability analysis of eigenvalues of the renormalisation group around
the fixed point. The philosophy is that these systems will ultimately
leave the vicinity of the fixed point.

In the second part of \cite{0205} (see also the last section of
\cite{Re1}) we undertook to sketch a rigorous renormalisation
framework for systems, lying on! the critical surface. Due to the
inherent long-range correlations, one must be extremely careful in
performing such an analysis. As such a rigorous analysis is both technically
demanding and a little bit tedious and incorporates a variety of
interesting mathematical side aspects like e.g. a singularity analysis
of distributions and pseudo differential operators, we decided to
separate this rather technical investigation off and give only a brief
discussion of one of the methods in this subsection, which we
exemplify with the help of the pair correlation function.

The general idea is it, to extract and isolate the characteristic
singular behavior of the correlation functions which is responsible
for the weak decay of correlation.  With $W^T(x)$ the truncated
two-point function, we, making the preceding analysis more rigorous,
assume the existence of a certain exponent, $\alpha$, so that ($x^2$
denoting the vector-norm squared) we can make the following
decomposition.
\begin{equation}G(x):=W^T(x)\cdot(1+x^2)^{(n-\alpha)/2}=const+F(x)\end{equation}
with a decaying (non-singular) $F$ which is assumed to be in $L^1$.
Fourier transformation then yields:
\begin{multline}R^{-2\gamma}\cdot\int
  W^T_2((x_1-x_2)+R(X_1-X_2))f(x_1/R)f(x_2/R)dx_1dx_2\\
=R^{-2\gamma}\cdot\int
G((x_1-x_2)+R(X_1-X_2))\cdot[1+((x_1-x_2)+R(X_1-X_2))^2]^{-(n-\alpha)/2}\cdot\\f(x_1/R)f(x_2/R)dx_1dx_2\\
=R^{-2\gamma}\cdot R^{2n-(n-\alpha)}\cdot\int dp\,\hat{G}(p)\cdot
e^{-iRp(X_1-X_2)}\cdot\\
\left[\int e^{-iRp(x_1-x_2)}(R^{-2}+((x_1-x_2)+(X_1-X_2))^2)^{-(n-\alpha)/2}f(x_1)f(x_2)dx_1dx_2\right]\end{multline}
where we made the substitution $x\to R\cdot x$.

As the support of $f$ is in principle arbitrary, we now assume it to
be contained in a sufficiently small ball around zero (or,
alternatively, $(X_1-X_2)$ sufficiently large so that
$(x_1-x_2)+(X_1-X_2)\neq 0$ for $x_i$ in the support of $f$).  With
\begin{equation}\hat{G}(p)=const\cdot\delta(p)+\hat{F}(p)\end{equation}
the leading part in the scaling limit $R\to\infty$ is the
$\delta$-term. Asymptotically we hence get for $R\to\infty$ (setting
$y:=x_1-x_2\;Y:=X_1-X_2$) the result, already conjectured in the
preceding subsection:
\begin{equation}R^{n+\alpha-2\gamma}\cdot const\cdot\int
  |y+Y|^{-(n-\alpha)}\cdot f\ast f(y)dy\end{equation}
with 
\begin{equation}f\ast f(y):=\int f(y+x_2)\cdot f(x_2)dx_2\end{equation}
and $y+Y\neq 0$ on $supp(f)$.

The reason why the contribution, coming from $\hat{F}(p)$, can be
neglected for $R\to\infty$ is the following: $f$ is assumed to be in
$\mcal{D}$; by assumption the prefactor never vanishes on the support
of $f(x_i)$. Hence the whole integrand in the expression in square
brackets is again in $\mcal{D}$ and therefore its Fourier transform,
$\hat{g}(p')$, is in $\mcal{S}$ (with $p':=Rp$), that is, of rapid
decrease. We can therefore perform the $R$-limit under the integral
and get a rapid vanishing of the corresponding contribution in $R$ for
 $R\to \infty$.
\begin{equation}\lim_{R\to\infty}R{-n}\cdot\int \hat{F}(p'/R)\cdot
  e^{-ip'Y}\cdot \hat{g}(p')d^np'=0 \end{equation}

As $f\ast f$ has again a compact support, we have that, choosing
\begin{equation}\gamma=(n+\alpha)/2\end{equation}
the limit correlation function behaves as $\sim
|X_1-X_2|^{-(n-\alpha)}$ as in the more heuristic analysis heuristic
analysis of the preceding subsection.

Along these lines, or choosing a slightly different method (see
\cite{0205}), one can proceed in the more difficult case of $l$-point
functions. In the course of this analysis an interesting phenomenon
does pop up which leads to some remarkable constraints as to the
consistency of the whole renormalisation picture. The critical
exponents are typically fixed by the assumed non-vanishing of the
scaled auto-correlation functions. On the other hand, the truncated $l$-point
functions may have a much more intricate cluster behavior (having, in
particular, a variety of decay channels). If one wants to go beyond
quasi-free limit theories, some higher truncated correlation functions
have to be non-vanishing in the scaling limit. For this to be the
case, there has to be some fine-tuning between their decay behavior
and the values of the critical exponents, which one got from the
two-point functions.
\section{\label{class}Some Remarks on Classical Statistical Systems}
In this section we want to briefly indicate how our framework can be
implemented in the regime of classical statistical mechanics. The
situation is more or less obvious in the class of spin- or
lattice-systems. The translation group is replaced by some discrete
lattice group. The Fourier vectors run through some Brillouin zone
instead of $\R^n$, while in coordinate space we employ the same kind
of smearing and averaging functions as in the continuous case, the
only difference being the replacement of integrals by sums.
For continuous classical KMS-systems, some more words are perhaps in
order (cf. e.g. \cite{Mermin}, \cite{Verboven}, \cite{Goldstein},
\cite{Zeitschr}, \cite{Wagner}). 

As infinitely extended phase space , $X$, we take the set of
sequences, $x$,
\begin{equation}x=(r_i,p_i)_{i=1}^{\infty}=(x_i)_{i=1}^{\infty}\end{equation}
of points 
\begin{equation}x_i=(r_i,p_i)\in \R^n\times\R^n\;,\;r_i\neq
  r_j\;\text{for}\; i\neq j  \end{equation}
having the \tit{local finiteness property}, i.e., the
number of points, $r_i$, occurring in $x$, is finite in each bounded set
, $V\subset \R^n$.

As local \tit{m-particle observables}, $A^{(m)}$, we take
\begin{equation}A^{(m)}(x):=\sum_{i_1<\cdots<i_m}f^{(m)}(x_{i_1},\ldots,x_{i_m})\end{equation}
$f$ from the class of smooth function with compact support in
coordinate space (the details are of course a matter of convenience).
\tit{Poisson brackets} can then be defined as usual and the local
finiteness property guarantees that the expression
\begin{equation}\{A,B\}(x):=\sum_{j=1}^{\infty}(\partial A/\partial
  r_j\cdot \partial B/\partial p_j-\partial A/\partial p_j\cdot
  \partial B/\partial r_j)    \end{equation}
is well-defined.

The thermodynamic equilibrium states are now probability measures on
the Borel-$\sigma$-field, defined on the phase space equipped with the
topology canonically induced by the class of observables. The
\tit{classical KMS-condition} we usually employ in the form ($A,B$
real):
\begin{equation}\langle\{A,B\}\rangle=\beta\langle B\{A,H\}\rangle  \end{equation}
$H$ being the Hamiltonian.

As in quantum statistical mechanics, we can define certain
distributional \tit{point fields} or \tit{densities} at, say,
coordinate $r$ over the phase space, like e.g.\\
\tit{particle density}:
\begin{equation}n_r(x):=\sum_i\delta(r-r_i)      \end{equation}
\tit{momentum density}:
\begin{equation}p_r(x):=\sum_i p_i\cdot\delta(r-r_i)     \end{equation}
energy density, stress tensor density etc. (see in particular
\cite{Zeitschr} and \cite{Wagner} where these notions have been
systematically employed).

The \tit{l-point distribution functions} can hence be expressed as
follows:
\begin{equation}\rho^{(l)}(r_1,\ldots,r_l):=\sum_{i_1<\cdots<i_l}\langle
\delta(r_1-r_{i_1})\cdots\delta(r_l-r_{i_l})\rangle      \end{equation}
$i_{\nu}$ running through the indices occurring in $x$. Ordinary
observables can be reconstructed by integrating these densities over
local test functions. For a one-particle observable we have for example:
\begin{equation}A_f:=\int a(r)\cdot f(r)d^nr    \end{equation}
with $a(r)$ a one-particle density and correspondingly for more
complex densities.

From these remarks one sees immediately, that the whole procedure, we
develop in the following, can be immediately transferred to the regime
of classical statistical mechanice without significant changes.
\section{\label{examples}A Class of Examples}
We argued that in the case of poor, that is, non-integrable
clustering, it appears to be mathematically more reasonable to perform
most of the necessary analysis in coordinate space, as the behavior in
Fourier space my be quite involved in the vicinity of
$(\omega,k)=(0,0)$.

The situation improves however if one has a more precise knowledge of
the form of correlation functions in Fourier space near $(0,0)$. We
note in passing that our approach is by no means restricted to the
case of critical systems. It does also apply to systems at zero
temperature or systems above or below a phase transition line. One may
have more precise information in Fourier space in various situations
like e.g. \tit{spontaneous symmetry breaking} (see \cite{cluster} and
\cite{Re1} and further literature given there) or for particular
correlation functions and/or commutators (so-called
\tit{sum-rules}). The relevant contribution in e.g. the 2-point
function can stem from sharp excitation branches or excitations,
having a finite lifetime, which is the typical situation in
interacting many-body systems. \\[0,3cm]
Remark: The assumptions in the following discussion can be
considerably weakened and are only made to cover a sufficiently
general and coherent class of models.\vspace{0.3cm} 

In order to better understand the effects of our general scaling
approach, we deal in this section with a fairly large class of
relatively manageable and simple models at non-zero temperature which
belong to the group of \tit{quasifree systems}. Note however that in
contrast to, say, relativistic quantum field theory, we have in
general no strong covariance properties. That is, even quasifree
systems are not completely uninteresting and supply us with a whole
bunch of useful model systems approximating important non-trivially
interacting systems. As this notion slightly varies from author to
author, we make the following assumptions.
\begin{assumption}Our class of models is assumed to have the following
  properties (in addition to the usual standing assumptions, we do not
  repeat here; see e.g. \cite{Bratteli2})\\
  The KMS-representations, $\pi_{\beta}$, of the quasi-local algebra
  $(\mcal{A},\alpha_{t,x})$ are assumed to be quasifree and faithful,
  that is
\begin{enumerate}
\item All n-point functions are products of 2-point functions.
\item $\pi_{\beta}(A)\neq 0$ if $A\neq 0$ in $\mcal{A}$.
\end{enumerate}
\end{assumption}
The second assumption seems to be physically reasonable (and can in
principle be weakened) as it avoids redundancies but need not! be
fulfilled in general. This situation occurs of course when the
original algebra has a non-trivial center and one studies
representations which are factors, in which central elements are
mapped onto c-numbers. Note that $\omega_{\beta}$, the KMS-state, is
always faithful (that is, \tit{separating}) in
$\pi_{\beta}(\mcal{A})''$ (the \tit{GNS-representation}), however this
need not be the case with respect to $\mcal{A}$ itself.

This point is relatively subtle from a more physical point of view and
not much seems to be known. There is a discussion in
\cite{Bratteli2},p.85ff.  which is based on the weak closure,
$\mcal{A}''$ of the original algebra. But, typically, an equilibrium
state is given via its local restrictions in form of Gibbs-states,
that is, it is naturally only defined on quasi-local elements of the
algebra and not on the weak closure. If $\mcal{A}$ is \tit{simple},
the representation is \tit{faithful}. (For an example of a
non-faithful representation see \cite{Bu2}). Note that the so-called
\tit{order parameters}, the non-vanishing of which usually signal the
occurrence of new phases, are typically global ``observables'' (for
example, meanvalues, not belonging to the algebra of quasi-local
observables) and are c-numbers in pure phases, i.e. factor states.

The above assumptions have both a simple technical consequence and a
consequence which is perhaps remarkable from a more physical point of view.
\begin{lemma}Under the assumptions being made the commutators in each
  representation, $\pi_{\beta}$, are c-numbers which do not depend on
  the KMS-state, that is, in contrast to the 2-point functions, they
  are state-independent.
\end{lemma}
Proof: i) The c-number property follows immediately from the vanishing
of all higher truncated correlation functions and is in fact
independent of the other assumptions. With
$span(\pi_{\beta}(\mcal{A})\Omega_{\beta})$ being dense in the
GNS-Hilbert space and
\begin{equation}(\Omega_{\beta},\pi_{\beta}(A)\cdot
  [\pi_{\beta}(B(x,t)),\pi_{\beta}(C)]\cdot
  \pi_{\beta}(A')\Omega_{\beta})\end{equation}
being a sum of 2-point functions, this expression can be shown to be equal to 
\begin{equation}(\Omega_{\beta},[\pi_{\beta}(B(x,t)),\pi_{\beta}(C)]\Omega_{\beta})\cdot (\Omega_{\beta},\pi_{\beta}(A)\cdot\pi_{\beta}(A')\Omega_{\beta})
  \end{equation}
ii) The faithfulness of $\pi_{\beta}$ implies (with
$C^{\beta}_{BC}(x,t)$ a function, which follows from i)) 
\begin{equation}C^{\beta}_{BC}(x,t)=[\pi_{\beta}(B(x,t)),\pi_{\beta}(C)]=\pi_{\beta}([B(x,t),C])=c_{BC}(x,t)=[B(x,t),C]      \end{equation}
with $c_{BC}(x,t)=C^{\beta}_{BC}(x,t)$ being \tit{independent} of the concrete
KMS-representation as
\begin{equation} \pi_{\beta}([B(x,t),C]-C^{\beta}_{BC}(x,t)\cdot\mbf{1})=0 \end{equation}\bewende

The physical relevance of the above observation is the following. With
\begin{equation}F^{\beta}_{AB}(x,t):=(\Omega_{\beta},\pi_{\beta}(A)(x,t)\cdot\pi_{\beta}(B)\Omega_{\beta})^T \end{equation}
and
\begin{equation}C^{\beta}_{AB}(x,t):=(\Omega_{\beta},[\pi_{\beta}(A)(x,t),\pi_{\beta}(B)]\Omega_{\beta}) \end{equation}
we have the general expression for the respective Fourier transforms
\begin{equation}\hat{F}^{\beta}_{AB}(\omega,k)=(1-\exp({-\beta\omega}))^{-1}\cdot\hat{C}^{\beta}_{AB}(\omega,k) \end{equation}

Usually, both $\hat{F}$ and $\hat{C}$ depend on the parameters,
fixing the KMS-state. Our assumptions guarantee that for our model
class the temperature dependence on the rhs is entirely concentrated
in the prefactor, $(1-\exp({-\beta\omega}))^{-1}$, that is, we have
\begin{koro}For our model class it holds 
\begin{equation}\hat{F}^{\beta}_{AB}(\omega,k)=(1-\exp({-\beta\omega}))^{-1}\cdot\hat{c}_{AB}(\omega,k) \end{equation}
with $\hat{c}_{AB}(\omega,k)$ temperature independent.
\end{koro}
As $\hat{c}_{AB}(\omega,k)$ is universal, it is typically simple to
calculate; use e.g. some ground state representation.

For the further analysis we choose $A,B$ selfadjoint and get for
$\hat{C}(\omega,k)$ (we supress the labels $A,B$):
\begin{equation}\hat{C}(\omega,k)=\hat{F}(\omega,k)-
  \overline{\hat{F}}(-\omega,-k)=(1-\exp({-\beta\omega}))\cdot F(\omega,k) \end{equation}
and hence
\begin{equation}Re\,\hat{F}(-\omega,-k)=\exp({-\beta\omega})\cdot Re\,\hat{F}(\omega,k)     \end{equation}
\begin{equation}Im\,\hat{F}(-\omega,-k)=- \exp({-\beta\omega})\cdot Im\,\hat{F}(\omega,k)     \end{equation}
thus clearly exhibiting the two-sidedness of the $(\omega,k)$-spectrum
in temperature states.

As, in contrast to the relativistic context (cf. e.g. \cite{Bu1} or
\cite{Bu2}), we have in general no strong covariance and/or
spectrum conditions for the 2-point functions, we have to make some
reasonable assumptions which are fulfilled in typical many-body
systems (for more details see \cite{cluster} and \cite{thirring}).   
\begin{assumption}We assume that the excitation spectrum of
  $\hat{F}(\omega,k)$ fulfills $\hat{F}(\omega,k)=\hat{F}(\omega,-k)$
  and contains a sharp excitation branch ($e(k)=e(|k|)$), describing
  stable quasi particles or collective excitations, with the remaining
  part being integrable and absolutely continuous around $(\omega,k)=(0,0)$. We denote the
  singular contribution by
\begin{equation}\hat{F}_{sing}(\omega,k):=J_+^{\beta}(k)\cdot\delta(\omega-(e(k)-\mu))+J_-^{\beta}(k)\cdot\delta(\omega+(e(k)-\mu))    \end{equation}
\end{assumption}
Remark: Note that in the translation invarinant case the above Fourier
transforms are measures!\vspace{0.3cm}

From the above relations we conclude that
\begin{equation}Re\,J_-^{\beta}(k)=\exp{(-\beta(e(k)-\mu))}\cdot Re\,J_+^{\beta}(k) \end{equation}
\begin{equation}Im \,J_-^{\beta}(k)=-\exp{(-\beta(e(k)-\mu))}\cdot Im \,J_+^{\beta}(k) \end{equation}
with $\mu$ the (temperature dependent; in case temperature and density
are chosen as independent parameters) \tit{chemical potential}. We
arrive at the following result:
\begin{lemma}We have
\begin{equation}J_-^{\beta}(k)=\exp{(-\beta(e(k)-\mu))}\cdot\overline{J_+^{\beta}(k)}\end{equation}
and
\begin{multline}\hat{C}_{sing}(\omega,k)=(1-\exp{(-\beta(e(k)-\mu))})\cdot
  J_+^{\beta}(k)\delta(\omega-(e(k)-\mu))\\-(1-\exp{(-\beta(e(k)-\mu))})\cdot\overline{J_+^{\beta}(k)}\delta(\omega+(e(k)-\mu))
\end{multline}
As $\hat{C}_{sing}(\omega,k)$ has to be independent of $\beta$ for our
class of models, we have furthermore
\begin{equation}J_+^{\beta}(k)=(1-\exp{(-\beta(e(k)-\mu))})^{-1}\cdot j(k) \end{equation}
\end{lemma}
Proof: This follows directly from the preceding formulas.\bewende
\vspace{0.3cm}

As our commutator function is universal, it should not contain the
typical singularities which show up in connection with phase
transitions and critical phenomena. As to this point we refer to the
discussion in e.g. \cite{cluster} and \cite{thirring}. These phenomena
are typically representation dependent. Therefore, on physical
grounds, the function $j(k)$ should be bounded near $k=0$ and $e(k)$
can be identified with the dispersion law of an \tit{elementary
excitation} which, in the non-relativistic context, for short-range
interactions, passes through zero for $k\to 0$.\\[0.3cm]
Remark: In \cite{cluster} we discussed various dispersion laws.
Frequently a simple power law behavior prevails.  \vspace{0.3cm}

We now apply our scaling procedure to the class of model systems
described above. In a first step we want to choose the scaling
exponent, $\gamma=\gamma_A$, in the expression
\begin{equation}A_R=R^{-\gamma}\cdot\int A(x+RX)\cdot
  f_R(x)d^nx\end{equation}
so that the corresponding autocorrelation function (remember the
standing assumption $\langle A\rangle=0$)
\begin{equation}\langle A_R(RX_1)\cdot A_R(RX_2)\rangle   \end{equation}
is both finite and non-vanishing in the limit $R\to\infty$.

Our above made observations or assumptions about the spectrum of the
2-point functions show that, provided we have a more detailed knowledge
of the system under discussion, we can, even in the case of long-range
correlations, perform the analysis in Fourier space getting
\begin{multline}\langle A_R(RX_1)\cdot A_R(RX_2)\rangle   =\\const\cdot
  R^{-2\gamma+n}\cdot\int\exp(-ik'(X_1-X_2))\cdot
  \hat{f}(k')\hat{f}(-k')\cdot\hat{F}_{AA}(\omega,k'/R)d\omega dk'\end{multline}
with $k':=Rk_1$ and $\hat{f}(k')\hat{f}(-k')=|\hat{f}(k')|^2$ for
$f(x)$ symmetric and real.

In the following we are concerned with the renormalisation of the
singular part of the spectral contribution as the absolutely
continuous part is (by assumption) harmless. For $\omega\geq 0$ we have
to consider the term
\begin{multline}\lim_{R\to\infty}I_R:=\\
\lim_{R\to\infty}
R^{-2(\gamma-n/2)}\cdot\int\exp(ik(X_1-X_2))\cdot(1-\exp(-\beta(e(k/R)-\mu)))^{-1}\cdot j(k/R)\cdot |\hat{f}(k)|^2d^nk\end{multline}

We do not intend to discuss the mathematically most general case but
rather concentrate on situations which are reasonable from a physical

point of view. 
\begin{assumption}We assume that in leading order $e(k)$ behaves like 
\begin{equation}e(k)\sim |k|^{\alpha}\quad\text{for}\quad |k|\to
  0\end{equation} 
with $\alpha>0$.
\end{assumption}
There is the possibility that $j(0)$ is finite but non-vanishing or
that $j(k)$ vanishes for $k\to 0$. We begin with the discussion of the
case of non-vanishing $j(0)$.\\[0.3cm]
I) \underline{$e(k)\sim |k|^{\alpha}$ near $k=0$, $j(k)$ continuous and $\neq 0$
in $k=0$}
\begin{ob}i) For $\mu\neq 0$ nothing peculiar happens and we are in the
  normal situation with $\gamma=n/2$.\\
ii) For $\mu=0$, the typical situation at or below the critical point,
we have the following behavior
\begin{equation}I_R\sim R^{-2(\gamma-(n+\alpha)/2)}\quad\text{for}\quad
R\to\infty\end{equation}
hence, the anomaleous scaling dimension is 
\begin{equation}\gamma=(n+\alpha)/2 \end{equation}
with $\alpha$ the exponent in the dispersion law of the sharp
elementary excitation mode.
\end{ob}
II) \underline{$j(k)$ vanishing in $k=0$}\\[0.3cm]
This situation is by no means entirely exceptional. Take for example
the time derivative at $t=0$ of the observable $A$. In the spectrum of
the autocorrelation function this leads to an additional prefactor,
$\omega^2$, in front of $\hat{F}_{AA}(\omega,k)$. In the singular
contribution, $J_+(k)$, this results in an additional factor,
$e(k)^2$, and hence in an additional contribution in the scaling exponent   
\begin{equation}\gamma_{\partial A}=n/2+\alpha/2-\alpha=\gamma_A-\alpha \end{equation}
\begin{ob}If one wants $\partial_t A$ to be a non-vanishing observable
  in the scaling limit, its scale dimension has to be chosen as
\begin{equation}\gamma_{\partial_t A}=\gamma_A-\alpha  \end{equation}
\end{ob}

Similar considerations have to be made for other functions of
elementary observables. If the spectrum is known qualitatively as in
our case, this can in fact be done in every concrete case. Note
furthermore that the temperature independence of the commutator is
technically convenient but not absolutely necessary. The same
conclusions do hold if the spectral weight along the sharp excitation
branch is temperature dependent. However, in that case we do not have
an apriori knowledge as to its precise form which may vary with
$\beta$. One can also treat the case of excitations having a
\tit{finite lifetime} (cf. \cite{cluster}). The excitation branch now
has a finite width and the calculations become even more model
dependent. On the other side we proved in \cite{thirring} that for
$\beta\neq 0$ sharp
excitation branches typically belong to elementary excitations having
no interaction with the rest of the system. 
\begin{bem} As to interesting consequences concerning the fate of
  commutators (i.e. the quantum nature) in the scaling limit see the
  discussion in subsection \ref{quantum}.
\end{bem}

\section{\label{rigorous}  Rigorous Results on the (Quantum) System in the Intermediate
  Regime and in the Scaling Limit} In this section we assume that the
theory exists in the scaling limit provided that the scaling exponents
have been appropriately chosen. Under this proviso we investigate its
algebraic and dynamical limit structure.
\subsection{\label{scales}The Description of the System at Varying Scales}
In algebraic statistical mechanics we describe a system with the help
of an observable algebra, $\mcal{A}$, a state, $\omega$, or
expectation functional, $<\circ>$, a time evolution,
$\alpha_t$. Frequently one also employs the $GNS$-Hilbert space
representation of the theory, introduced by Gelfand, Naimark, Segal
(see e.g. \cite{Bratteli1}). We already gave a brief discussion of
these points in \cite{Re1}. But as the approach of the scaling limit
is quite subtle both physically and mathematically, we would like to
give a more complete discussion of some of the topics in the following.

We begin with fixing the notation and introducing some technical and
conceptual tools. Expectation values of elements of the underlying
observable algebra, $\mcal{A}$, at scale ``$0$'', are given by
\begin{equation}\omega(A(1)\cdots A(l))=:\langle A(1)\cdots
  A(l)\rangle\end{equation}
where different indices may denote different
elements of the algebra,
different times etc. The dynamics is denoted by
\begin{equation}\alpha_t(A)=A(t)\;\text{or}\;A_t\;,\;t\in\R \end{equation}
space translations by
\begin{equation}\alpha_x(A)=A(x)\;\text{or}\;A_x\;,\;x\in\R^n \end{equation}
\begin{equation}\alpha_{t,x}(A)=A(t,x)   \end{equation}

Given such a structure, we can construct a corresponding Hilbert space
representation (for convenience, we use the same symbols for the
elements of the original algebra and their representations in the
$GNS$-representation).
\begin{equation}\omega\to\Omega\;,\;\omega(A(1)\cdots
  A(l))=(\Omega|A(1)\cdots A(l)\Omega)_{GNS
}   \end{equation}
\begin{equation}\alpha_t\to U_t\;,\;\text{with}\;\alpha_t(A)\to
  U_t\cdot A\cdot U_{-t}  \end{equation}

The averaged or renormalized observables, $A\to A_R$, at scale $R$ are
a subset of elements contained in the original algebra, $\mcal{A}$. We
denote the subalgebra, generated by these elements, by $\mcal{A}_R$
with $\mcal{A}_R\subset\mcal{A}$. We can decide to forget the finer
algebra, $\mcal{A}$, and define the \tit{algebra on scale} $R$:
\begin{defi}We define the system on scale $R$ by
\begin{equation}\omega^{(R)}(A^{(R)}):=\omega(A_R) \end{equation}
\begin{equation}\alpha_t^{(R)}(A^{(R)}):=(\alpha_t(A))^{(R)} \end{equation}
\begin{equation}\alpha_X^{(R)}(A^{(R)}):=(A(RX))^{(R)}    \end{equation}
more specifically, we define the objects on the lhs implicitly (via
the \tit{GNS-reconstruction}) by the following correspondence
\begin{equation}\langle A^{(R)}(t_1,X_1)\cdots
  A^{(R)}(t_l,X_l)\rangle_{(R)}:=\langle A_R(t_1,RX_1)\cdots A_R(t_l,RX_l)\rangle  \end{equation}
\end{defi}
Remark: Note the different treatment of time and
space-translations. We will come back to this point (which has
remarkable physical consequences) below in connection with
\tit{critical slowing down}.
\begin{satz}
  From the above we see that on each scale we have a new theory,
  $\mcal{S}^{(R)}$ ($\mcal{S}$ standing for ``system''), which we get
  by reconstruction from the above hierarchy of correlation functions,
  in particular, a new, non-isomorphic algebra, $\mcal{A}^{(R)}$, and
  a corresponding $GNS$-Hilbert space representation. We emphasize
  that the coarse-grained dynamics is also physically different
  (despite the seeming similarity of the expressions on both sides of
  the above definitions).

If the scaling limit does exist, we have, by the same token, a scaling
limit system denoted by
\begin{equation}\mcal{S}^{\infty}=(\omega^{\infty},\mcal{A}^{\infty},\alpha_t^{\infty},\alpha_X^{\infty})\end{equation}
with
\begin{equation}\langle A^{\infty}(t_1,X_1)\cdots
  A^{\infty}(t_l,X_l)\rangle=\lim_{R\to\infty}\langle
  A_R(t_1,RX_1)\cdots A_R(t_l,RX_l)\rangle     \end{equation}
\end{satz}
The proof is more or less obvious from what we have said
above.\bewende      \\[0.3cm]
\begin{koro}We generally assume that $\alpha_t$ is strongly continous
  on $\mcal{A}$. By the above identification process we can
  immediately infer that both $\alpha_t^{(R)}$ and $\alpha_t^{\infty}$
  are also strongly continuous on the corresponding algebras,
  $\mcal{A}^{(R)},\mcal{A}^{\infty}$. By the same token, we can infer
  that $\omega^{(R)}$ and $\omega^{\infty}$ are  $KMS$-states at the
  same inverse temperature $\beta$.
\end{koro}
\begin{bem}Strong continuity can be generally achieved by going over
  to smoothed observables, i.e., by averaging the observables with
  smooth functions of, say, compact support in the time variable.
\end{bem}
Proof of Corollary: Note that the original time evolution ``commutes''
with the scale transformation in the sense described above. This
yields the mentioned result for all finite $R$. We have in particular
that for suitable elements (for the technical details see
\cite{Bratteli2})
\begin{equation}\langle B^{(R)}(t)\cdot A^{(R)}\rangle_{(R)}=\langle A^{(R)}\cdot B^{(R)}(t+i\beta)\rangle_{(R)} \end{equation}
and there exists an analytic function, $F_{AB}^{(R)}$(z), in the strip
$\{z=t+i\tau,\;0<\tau<\beta\}$ with continuous boundary values at
$\tau=0,\beta$:
\begin{equation}F_{AB}^{(R)}(t)= \langle A^{(R)}\cdot B^{(R)}(t)\rangle_{(R)}\;,\; F^{(R)}_{AB}(t+i\beta)= \langle A^{(R)}\cdot B^{(R)}(t+i\beta)\rangle_{(R)}    \end{equation}
This is equivalent to the following equation (cf. \cite{Bratteli2}):
\begin{equation}\int\omega^{(R)}(A^{(R)}\cdot B^{(R)}(t))\cdot f(t)dt=
  \int\omega^{(R)}(B^{(R)}(t)\cdot A^{(R)})\cdot f(t+i\beta)dt    \end{equation}
for $\hat{f}\in \mcal{D}$. As $f(t+i\beta)$ is of strong decrease in
$t$ the limit $R\to\infty$ can be performed under the integral and we
get the same relation in the scaling limit. The above mentioned
equivalence of this property with the $KMS$-condition shows that the
limit state is again $KMS$. This proves the statement.\bewende   \\[0.3cm]
Remarks: i) Note what we have already said in \cite{Re1}. One reason for
the non-equivalence of the algebras on different scales stems from the
observation that, in general, 
\begin{equation}A_R\cdot B_R\neq (A\cdot B)_R\end{equation}
Furthermore, in the scaling limit, many different observables of
$\mcal{A}$ converge to the same limit point, for example, all finite
translates of a fixed observable.\\
ii) A corresonding result in a slightly different context was also
proved in \cite{Bu}.
\subsection{The Scaling Limit Theory as a Quantum Field Theory}
We have seen in sect. 2.3 that the scaling limit of the correlation
functions for the block spin observables is not fully scale invariant
but only asymptotically so (while the short range details of the
original microscopic correlations, encoded in the function
$F(x_1-x_2)$, have been integrated out, there remains an integrated
effect of the initial block-function, $f(x)$ ).

This observation runs a little bit contrary to the general folklore,
in which the various limit procedures are frequently interchanged and
identified without full justification. We will exhibit the true
connections between the various expressions in the following.

With $f(x)$ now being a \tit{general} test function of e.g. compact
support, we have from sect. 2.3, making now the dependence on $f$
explicit
\begin{equation}\lim_{R\to\infty}\langle A_{R,f}(RX_1)\cdot  A_{R,f}(RX_2)\rangle=const\cdot\int |y+Y|^{-(n-\alpha)}\cdot f\ast f(y)dy    \end{equation}
with
\begin{equation}A_{R,f}(RX)=R^{-(n+\alpha)/2}\cdot\int A(RX+x)\cdot f(x/R)dx   \end{equation}
We now rewrite the limit correlation function as
\begin{equation}\langle A^{\infty}_f(X_1)\cdot
  A^{\infty}_f(X_2)\rangle=\int \langle \hat{A}^{\infty}(x_1+X_1)\cdot
  \hat{A}^{\infty}(x_2+X_2)\rangle\cdot f(x_1)f(x_2)dx_1dx_2  \end{equation}
that is, we identify 
\begin{equation}A_f^{\infty}(X)=\int \hat{A}^{\infty}(x+X)\cdot f(x)dx   \end{equation}
with $ \hat{A}^{\infty}(x)$ now having rather the character of a \tit{field}
or operator valued distribution.

We have that
\begin{equation}\langle \hat{A}^{\infty}(X_1)\cdot
  \hat{A}^{\infty}(X_2)\rangle=:W^{\infty}(X_1-X_2)= const\cdot |X_1-X_2|^{-(n-\alpha)}
   \end{equation}
Corresponding results would hold for the higher correlation functions,
that is, we arrive at
\begin{conclusion}In contrast to the block observables,
  $A_f^{\infty}$, the field, $\hat{A}^{\infty}(x)$, displays the full scale
  invariance.
\end{conclusion}

The field, $\hat{A}^{\infty}(x)$, can, on the other hand, be directly
constructed by means of a related limit procedure, which is however
\tit{not} of block variable type. We start instead with the
\tit{unsmeared} observables and take the scaling limit, $R\to\infty$
\begin{equation}\lim_R\langle\hat{A}_R(RX_1)\cdot
  \hat{A}_R(RX_2)\rangle\;\text{with}\;\hat{A}_R(RX):=R^{(n-\gamma)}\cdot A(RX)     \end{equation}
and $n-\gamma=(n-\alpha)/2$.\\[0.3cm]
Remark: The extra scaling factor, $R^n$, replaces the missing
integration over the test function $f_R$, the support of which increases
like $R^n$.\vspace{0.3cm}

Performing the same calculations, we see that the above limit is equal to\\
$\langle A^{\infty}(X_1)\cdot A^{\infty}(X_2)\rangle$. We hence have
\begin{conclusion}The fully scale invariant limit theory is achieved
  by taking the limits
\begin{equation}\lim_R\langle\hat{A}_R(RX_1)\cdots \hat{A}_R(RX_l)\rangle=:W^{\infty}(X_1,\ldots,X_l)   \end{equation}
The same construction holds of course for the intermediate scales; we
define $\hat{A}^{(R)}(X)$ by the following identification
\begin{equation}\langle\hat{A}^{(R)}(X_1)\cdots
  \hat{A}^{(R)}(X_l)\rangle_{(R)}:=R^{l(n-\gamma)}\cdot\langle
  A(RX_1)\cdots A(RX_l)\rangle\end{equation}
and have for the observables, $A_f^{(R)}$, defined above
\begin{equation}A_f^{(R)}(X)=\int \hat{A}^{(R)}(X+x)f(x)dx \end{equation}
(which can e.g. be checked by direct calculation). 
\end{conclusion}
\subsection{\label{quantum} The (Non)-Quantum Character in the Scaling Limit}
In subsection B of section 3 of \cite{Re1}, we already discussed the
limiting behavior of commutators of scaled observables. In the regime
of \tit{normal} scaling, that is, scale dimension $\gamma=n/2$, we
found that commutators are non-vanishing in the generic case in the
limit. This means that in general the resulting limit theory is
\tit{non-abelian} (but \tit{quasi-free}!). Perhaps a little bit
surprisingly, the situation changes at the \tit{critical point}, where
the scale-dimensions are, typically, greater than $n/2$ for at least
some observables!

We make the same observation as Sewell in \cite{Sewell}, namely,
commutators of certain ``critical'' observables vanish in the scaling
limit, i.e., the corresponding limit observables are loosing (at
least) part of
their quantum character  .\\
[0.3cm] Remark: We think that the observation that fluctuations and
critical behavior at the critical point are typically of a
\tit{thermal} and not of a \tit{quantum} nature, does somehow belong
to the general folklore in the field of critical phenomena, but we are
not aware at the moment that this fact has been \tit{widely} discussed
in the literature in greater rigor. Some remarks can e.g. be
found in connection with so-called (\tit{temperature-zero})
\tit{quantum phase transitions} in \cite{Sachev} or \cite{Physik} and
further references given there.

On the other hand, related phenomena were observed in the context of
\tit{spontaneous symmetry breaking} in sect. 6 of \cite{Re1} and for
certain models by Verbeure et al in \cite{Verbeure}. A careful
analysis of the behavior of commutators in a slightly different
context can also be found in \cite{Zagrebnov}.
\\[0.3cm]
We have the following result.
\begin{satz}Let $A,B$ be strictly localized observables with
  $\gamma_A+\gamma_B>n$. We then have
\begin{equation}\lim_R\|[A_R,B_R]\|=0 \end{equation}
\end{satz}
Proof: With $\gamma_A+\gamma_B>n$ we have
\begin{multline}\|[A_R,B_R]\|\leq
  R^{-(\gamma_A+\gamma_B)}\cdot\int\|[A(x_1),B(x_2)]\|\cdot
  f(x_1/R)f(x_2/R)dx_1dx_2\\
=R^{-(\gamma_A+\gamma_B)}\cdot\int\|[A,B(y)]\|\cdot
f(x_1/R)f((x_1+y)/R)dx_1dy
\end{multline}
By assumption $A,B$ have bounded supports, $V_A,V_B\subset \R^n$ so that
\begin{equation}[A,B(x)]=0\;\text{for}\;V_B+x\cap V_A=\emptyset \end{equation}
From the support assumption we immediately infer that the above double
integral is actually a single integral as the commutator on the rhs
vanishes outside a set, $S$, of finite diameter. We get
\begin{multline}\lim_R \|[A_R,B_R]\|\leq const'\cdot
  R^{-(\gamma_A+\gamma_B)}\int\chi_S(y)\cdot
  f(x_1/R)f((x_1+y/R))dx_1dy\\
=const' R^{-(\gamma_A+\gamma_B)}\cdot R^n\int\chi_S(y)\cdot f(x'_1)f(x'_1+y/R)dx'_1dy \leq const\cdot\lim_R
  R^{n-(\gamma_A+\gamma_B)}=0 \end{multline}
as $\gamma_A+\gamma_B>n$ by assumption.\bewende

\begin{koro}We arrive at the same result if $A,B$ are not strictly
  localized but fulfill a norm estimate of the form
\begin{equation}\|[A,B(y)]\|=:F(y)\in L^1(\R^n) \end{equation}
\end{koro}
Proof: We have
\begin{multline}R^{-(\gamma_A+\gamma_B)}\cdot\int F(y)\cdot f(x_1/R)f((x_1+y)/R)dx_1dy=\\
R^{-(\gamma_A+\gamma_B)}\cdot R^{2n}\cdot\int
\hat{F}(p)\hat{f}(Rp)\cdot\hat{f}(-Rp)dp\\
=R^{-(\gamma_A+\gamma_B)}\cdot R^n\cdot\int\hat{F}(p/R)\hat{f}(p)\cdot\hat{f}(-p)dp  
\end{multline}
We can again perform the $R$-limit under the integral and get the
limit expression
\begin{equation}R^{n-(\gamma_A+\gamma_B)}\cdot\hat{F}(0)\cdot\int
  \hat{f}(p)\cdot\hat{f}(-p)dp\to 0   \end{equation}
for $R\to\infty$.\bewende

A simple example where different renormalisation exponents naturally
arise is the following. Take a limit observable, $A^{\infty}(X)$, and
consider its spatial derivative, $\partial_X A^{\infty}(X)$. Then we
have in a slightly sloppy notation (the limit being taken in the sense, described above):
\begin{multline}\partial_X A^{\infty}(X)=\lim_R
  \partial_X(R^{-\gamma_A}\cdot\int A(x+RX)\cdot f_R(x)d^nx)\\
=\lim_R (R^{(-\gamma_A+1)}\cdot\int (\partial_x A)(x+RX)\cdot f_R(x)d^nx)
\end{multline}
That is, $\partial_xA=i[P,A]$ has to be scaled with a different scale
exponent. Physically, this can be understood as follows. With
$f_R(x)=f(|x|/R)$ simulating the integration over a ball with radius
$R$, a partial integration in the above formula shifts the
$\partial_x$ to the test function, $f_R(x)$. As $\partial_x f_R(x)=0$
in the interior of the ball, the averaging goes roughly only over the
sphere of radius $R$ instead of the full ball. This has to be
compensated by a weaker renormalisation.

Another result in this direction can be found in \cite{Re1} sect.6, in
connection with the canonical Goldstone pair in the context of
spontaneous symmetry breaking.

Further possible candidates are the time derivatives of observables
as, for example, in $\langle \dot{A}\dot{A}\rangle$. Fourier
transformation yields an additional prefactor, $\omega^2$ in the
spectral weight, $\hat{F}_{AA}(\omega,k)$. The $KMS$-condition leads
to another constraint:
\begin{equation}\hat{F}_{AB}(\omega,k)=(1-e^{-\beta\omega})^{-1}\cdot
  \hat{C}_{[A,B]}(\omega,k)\end{equation}
A combination of such properties shows, that in the scaling limit, the
vicinity of $(\omega,k)=(0,0)$ is important (see the discussion in
section \ref{examples}). 

From covariance properties (as e.g. in models of relativistic quantum
field theory) one can infer additional information about certain
characteristics of the energy-momentum spectrum. For arbitrary
models of non-relativistic many-body theory, however, the situation is less
generic and typically model dependent.
\\[0.3cm]
Remark: We had several discussions with D.Buchholz about this point,
which are gratefully acknowledged. This applies also to the following
subsection.

\subsection{ \label{Slowing} The Nature of the Limit Time Evolution and the Phenomenon
  of Critical Slowing Down} We argued above that the appropriate
choice of the respective scaling dimensions of the observables under
discussion is a subtle point and perhaps, to some extent, even a
matter of convenience. After all, one may have some freedom in the
choice of the subset of observables which is to survive the
renormalisation process.

We will not give a complete analysis of all possibilities in the
following but rather emphasize one, as we think, particularly
remarkable phenomenon, namely, the phenomenon of \tit{critical slowing
  down}. As in the preceding discussion, we choose two observables,
$A,B$, with $\gamma_A+\gamma_B>n$, implying that the limit commutator
vanishes. We assume this also to hold for non-equal times (which
follows from a $L_1$-cluster condition as in the preceding
subsection), at least on the level of two-point functions, i.e.
\begin{equation}\langle [A^{\infty},B^{\infty}(t)]\rangle_{\infty}=0  \end{equation}
and get the following theorem:
\begin{satz}Under the assumptions being made, we have
\begin{equation}\langle A^{\infty}\cdot B^{\infty}(t)\rangle_{\infty}=
const \;\text{for all}\; t\in \R\end{equation}
\end{satz}
Proof: As the limit state is again a $KMS$-state, the vanishing of the
above commutator implies that the analytic function,
$F^{\infty}_{AB}(z)$, fulfills
\begin{equation}F^{\infty}_{AB}(t)=F^{\infty}_{AB}(t+i\beta)     \end{equation}
for all $t$. $F^{\infty}_{AB}(z)$ can hence be analytically continued
to the whole $\C$-plane and is, furthermore, a globally bounded analytic
function, hence a constant by standard reasoning.\bewende

We see that the subclass of limit observables, which has vanishing
limit commutators (see the preceding subsection), has, by the same
token, time independent limit correlation functions. As these
pair-correlation functions are usually connected with characteristic
observable properties of the system (generalized suszebtibilities,
transport coefficients etc.), this has remarkable physical
consequences. The corresponding phenomenon is called \tit{critical
  slowing down}. For a review of the physical phenomena see e.g.
\cite{Halperin}. In physical terms, the phenomenon can be understood
as follows.

In the critical regime, the patches of strongly correlated degrees of
freedom become very large and extend practically over all scales. That
is, a reorientation of such clusters or a response to external
perturbations takes, if viewed on the microscopic time scale, a very
long time. In the scaling limit this time scale goes to infinity. If
one wants to see observable dynamical effects on the macroscopic level
one must hence scale the time variable also. For the unscaled time we
have in the limit $R\to\infty$:
\begin{equation}d/dt\langle A^{\infty}\cdot
  B^{\infty}(t)\rangle_{\infty}=0\end{equation}
This is the same as
\begin{equation}\label{formula}  \langle A^{\infty}\cdot[H_{\infty},B^{\infty}(t)]
  \rangle_{\infty}=\lim_R \langle A_R\cdot[H,B_R(t)]
  \rangle=\lim_R d/dt\langle A_R\cdot  B_R(t)\rangle
 \end{equation}
(cf. subsection \ref{scales}, $H$ is the microscopic Hamiltonian).

What one now has to do is obvious. We have to compensate the vanishing
of the above expression in the limit by inserting an appropriate scale
factor in the time coordinate. Instead of $B(t)$ we take
$B(R^{\delta}\cdot t)$ with $\delta$ so chosen that the limit
expression is non-vanishing. Note that differentiation with respect to
$t$ now yields an explicit prefactor $R^{\delta}$. This fixes the
\tit{macroscopic} time scale, $t_m$, for these processes. We define
\begin{equation}\langle A^{\infty}\cdot B^{\infty}(t_m)\rangle_{\infty}:=
\lim_R\langle A_R\cdot  B_R(R^{\delta}\cdot t_m)\rangle
\end{equation}
Physically the effect can be understood by inspecting the
middle part of equation (\ref{formula}). The support of $B_R(t)$
spreads with time. This spread is more pronounced if we take
$R^{\delta}\cdot t$ instead of $t$. By the same token the overlap with
the Hamiltonian (which is basically translation invariant) increases
with $R\to\infty$ while $t$ is kept fixed, thus yielding the
non-vanishing limit.

It may happen that other observables may evolve on different macroscopic
time scales so that the construction of a coherent common macroscopic limit
time evolution may not be straightforward. Such more detailed questions have
to be separately studied for the various model classes. As we have
studied a concrete model class in section \ref{examples}, we can make
much more precise statements if we have some information about the
energy-momentum spectrum in the vicinity of $(\omega,k)=(0,0)$. In
that section we arrived at the following results:
\begin{equation}\hat{F}^{\beta}_{AB}(\omega,k)=(1-\exp({-\beta\omega}))^{-1}\cdot\hat{c}_{AB}(\omega,k)
\end{equation}
and
\begin{equation}\hat{F}_{sing}(\omega,k):=J_+^{\beta}(k)\cdot\delta(\omega-(e(k)-\mu))+J_-^{\beta}(k)\cdot\delta(\omega+(e(k)-\mu))    \end{equation}
with
\begin{equation}J_+^{\beta}(k)=(1-\exp{(-\beta(e(k)-\mu))})^{-1}\cdot j(k) \end{equation}
and
\begin{equation}J_-^{\beta}(k)=\exp{(-\beta(e(k)-\mu))}\cdot\overline{J_+^{\beta}(k)}\end{equation}

For e.g.
\begin{equation}\langle A_R(RX_1,t)\cdot A_R(RX_2,0)\rangle     \end{equation}
we have to study expressions like
\begin{equation}R^{-2(\gamma_A-n/2)}\cdot\int e^{i(e(k/R)-\mu)\cdot
    t}\cdot (1-\exp(-\beta(e(k/R)-\mu)))^{-1}\cdot j(k/R)\cdot |\hat{f}|^2d^nk     \end{equation}
Again, the situation is normal for $\mu\neq 0$ but becomes singular for
$\mu=0$ (cf. section \ref{examples}).

We concentrate on the case $e(k)\sim |k|^{\alpha}\quad\text{for}\quad
|k|\to 0$ and $j(k)$ continuous and $\neq 0$ in $k=0$
(cf. section\ref{examples} assumption \ref{examples}.6). In order to
have a non-vanishing limit correlation function we have to choose
\begin{equation}\gamma_A=(n+\alpha)/2 \end{equation}
In the case where $j(k)$ vanishes polynomially in $k=0$ we have to
make a corresponding choice, as has been described in the mentioned
section.

If, furthermore, we want to have a non-trivial time evolution in the
limit $R\to\infty$, we have to scale the \tit{microscopic} time like
\begin{equation}t=R^{\alpha}\cdot\tau\quad\text{so that}\quad
  |k/R|^{\alpha}\cdot t=|k|^{\alpha}\cdot\tau    \end{equation}
\begin{conclusion}If we are in the situation, described in section
  \ref{examples}, having a singular spectral contributution with
  quasi-particle-like dispersion law $e(k)\sim |k|^{\alpha}$ near
  $k=0$, we have to scale the microscopic time, $t$, like
  $t=R^{\alpha}\cdot\tau$, in order to arrive at a non-trivial limit
  time evolution in the variable $\tau$.
\end{conclusion}


Acknowledgement: Several discussions with D.Buchholz are greatefully
acknowledged (see also the remark at the end of subsection 5.3).

\end{document}